\documentclass[11pt]{article}
\usepackage{moriond,epsfig,amsmath,amssymb}

\def\be{\begin{equation}}
\def\ee{\end{equation}}
\def\bea{\begin{eqnarray}}
\def\eea{\end{eqnarray}}

\newcommand{\fig}[1]{fig.~\ref{fig:#1}}

\newcommand{\eq}[1]{eq.~(\ref{eq:#1})}

\newcommand{\TeV}{\,\mathrm{TeV}}
\newcommand{\GeV}{\,\mathrm{GeV}}
\newcommand{\MeV}{\,\mathrm{MeV}}

\newcommand{\eV}{\,\mathrm{eV}}

\newcommand{\erg}{\,\mathrm{erg}}

\newcommand{\YLe}{Y_{L_e}}

\newcommand{\mmR}{m^2\! R}
\newcommand{\nue}{\nu_e}
\newcommand{\nueb}{\bar\nu_e}
\newcommand{\num}{\nu_\mu}

\newcommand{\nut}{\nu_\tau}

\newcommand{\nB}{n_{B}}

\newcommand{\GF}{G_{\text{F}}}
\newcommand{\SN}{SN 1987A}
\newcommand{\km}{\,\mathrm{km}}

\newcommand{\gcm}{\,\mathrm{g}\,\mathrm{cm}^{-3}}
\newcommand{\Msun}{M_{\odot}}

\newcommand{\Mf}{M_*}
\newcommand{\meff}{m_{\text{eff}}}

\begin{document}
\vspace*{4cm}
\title{NEUTRINOS IN EXTRA DIMENSIONS AND SUPERNOVAE}

\author{ M. CIRELLI }

\address{Scuola Normale Superiore and INFN,\\
piazza dei Cavalieri 7, I-56126, Pisa, Italy}

\maketitle\abstracts{
Models with extra space dimensions naturally provide for the existence of fermions that propagate in them. 
These are seen in 4D as an infinite tower of sterile neutrinos possibly mixed with the Standard Model ones.
We consider the effect of such a mixing in the context of core collapse supernova physics.
We show that the potentially dramatic modifications to the supernova evolution (commonly believed to set very strong bounds on the parameters of the extra dimensions) are prevented by a mechanism of feedback, so that much weaker bounds need to be imposed.
Nevertheless, the supernova core evolution is significantly modified.  
We discuss the compatibility with the \SN\ signal and we analyse the distinctive signatures of the neutrino signal on Earth.}

\section{Introduction}\label{sec:introduction}
The success of supernova (SN) models, confirmed by observation, often implies strong constraints on several kinds of new physics that could affect the supernova evolution too drastically; in the case of the existence of extra dimensions, and of neutrinos propagating in them, those limits seem to be very stringent if taken at face value.
On another hand, even within the bounds, extra dimensions can modify SN evolution in some peculiar way, so to reveal their presence; and they could even have beneficial consequences on the flaws that SN models still present.
This twofold connection is a strong motivation for the combined study of supernova evolution and neutrinos in extra dimensions.

In this perspective, we will mainly address two questions: What are the actual limits on the parameters of the extra dimensions set by SN evolution? What will be the signatures of the presence of extra dimensions in the neutrino signal from a SN detectable on Earth?

The discussion is based on ref. \cite{paper1} and ref. \cite{paper2}. 

\section{The supernova game}\label{sec:supernova}
For our purposes, it will be sufficient to sketch the standard picture of the supernova phenomenon$\,$\cite{SN} in terms of three main phases: (i) the gravitational collapse, (ii) the (delayed) explosion and (iii) the cooling phase.\\
(i) The collapse begins when the iron core of the star, accreting matter from the outer layers but no more capable of nuclear burning, reaches the Chandrasekhar limit: the electrons cannot compensate the gravitational pressure any more.
During a fraction of a second the radius of the core gets reduced and its density increased by several orders of magnitude ($10^{9} \gcm \rightarrow \textrm{few} \cdot 10^{14} \gcm$). 
The capture reaction $p e^- \rightarrow n \nue$ leads to the neutronization of matter and produces a large number of very energetic electron neutrinos, that remain trapped when densities of order $10^{12} \gcm$ are reached.\\
(ii) The collapse abruptely stops when nuclear densities ($\sim 3 \cdot 10^{14} \gcm$) are reached: the falling material bounces on the surface of the inner core and turns the implosion of the core in an explosion of the outer layers. 
However, the outward propagating shock wave loses energy on the way, slows down and would eventually recollapse.
In the meantime, the cooling phase is beginning: the neutrinos contained in the inner core start to diffuse out.
They hit and push the stalling matter from below so that their energy deposition is believed to be essential for the actual explosion (delayed explosion picture).
Unfortunately, the current computer simulations that include standard physics and assume spherical symmetry do not generally succeed in reproducing an explosion.
It could simply be due to the excessive simplifications adopted or it could even be a hint for the need of new physics.\\
(iii) In the following, we will mainly play with the cooling phase, so let us focus on this stage a bit more closely.
The game consists in the diffusion of the very energetic ($E \sim 100 \MeV$) neutrinos out of the dense and hot inner core of the star (with mass $\sim \Msun \simeq 2\cdot 10^{33}\,\mathrm{g}$, radius $\sim 10 \km$, density $\sim \textrm{few} \cdot 10^{14} \gcm$).
The beta reactions effectively act as a continuous pumping of energy and lepton number from the core matter into the neutrinos, that carry them away.
The evolution can be completely described in terms of a few dynamic variables, to be followed in the core during the evolution: the temperature $T$, the matter density $\rho$ and the leptonic fractions $Y_{L_{e,\mu,\tau}} = Y_{\nu_{e,\mu,\tau}}+Y_{e,\mu,\tau}$, where $Y_x$ is the net number fraction per baryon of the species $x$: $Y_x=(n_x - n_{\bar x})/\nB$.
At the beginning, $T \sim 10-40 \MeV$ and $\YLe \sim 0.35$ (with some characteristic initial profile produced during the collapse phase). 
In such conditions $\nue$ are highly degenerate, with a large chemical potential. 
On the contrary, $Y_{L_{\mu,\tau}} = 0$, since these flavours are produced in pairs. 
Then neutrinos carry out of the core almost all the energy and the (electron) lepton number and leave, at the end of the process, a cold, deleptonized proto-neutron star.

\subsection{The energy loss constraint}

Some robust key features of the picture above can be highlighted:
first, the total energy emitted in neutrinos roughly corresponds to the gravitational energy of the progenitor star (estimated in $\sim \textrm{few} \cdot 10^{53} \erg$ for a typical star); 
second, the overall timescale of neutrino emission is predicted to be a few tens of seconds, since it is determined by the conditions of neutrino trapping.

Remarkably enough, these key features are confirmed (despite the low statistics) by \SN\,, the single supernova event in which we could detect the neutrino (namely $\nueb$) signal so far.\cite{1987a}
As a consequence, any loss in a channel that is alternative to neutrinos must neither drain a too large portion of the total energy nor shorten the neutrino emission too much.

This {\it energy loss constraint}, even though so simple, proves to be a stringent one for several kinds of modification that one is willing to introduce in the standard supernova evolution.

\section{The extra dimensional playground}\label{sec:playground}
\subsection{Why (and which) extra dimensions}
The introduction of large extra dimensions aims at solving the hierarchy problem (i.e. the hierarchy between the Planck scale $M_{Pl} \simeq 10^{19} \GeV$ and the electroweak scale $M_{EW} \sim \textrm{few} \cdot 100 \GeV$) simply removing the need for the Planck scale itself. 
Indeed, the only fundamental scale is set at $\Mf \sim \TeV$; gravitons are allowed to propagate in $\delta$ extra dimensions compactified on circles of radius $R_i$ so that the hugeness of the Planck scale is produced by $M_{Pl}^2=\prod_{i=1}^{\delta} (2 \pi R_i M_*) ~M_*^2$. 
The Standard Model fields are confined on a 4D brane.\cite{ED}

In the following we consider the single {\it largest} compact extra dimension, of radius $R$, on which the only request comes from direct tests of the Newton's law at small distances:
\be
\label{eq:Rlimits}
1/R \gtrsim 10^{-3} \eV \quad \textrm{i.e.} \quad R \lesssim 100 \mu m \ .
\ee

\subsection{Why fermions in extra dimensions}
On top of the scenario described above, a fundamental observation is that it is very natural to allow also for the propagation of {\it fermionic fields in the extra dimension}.\cite{nuinED1}
Indeed, for instance, once the existence of the extra dimension has been postulated, any field which is sterile under the SM gauge group (not only the graviton) has no good reason to be confined on the brane; the most natural candidate is the right handed neutrino, a fermion.
Also, if extra dimensions are inspired and implied by string/M theory, then several scalars are inevitably present in the bulk (e.g. the moduli that fix the internal radii) and so are their fermionic superpartners, since SuperSymmetry is also part of string theory.
With an obvious extension of terminology, we call these fermionic fields `bulk neutrinos' or `neutrinos in extra dimensions'. 

\subsection{A specific example and the general case}
Let us first discuss the simplest paradigm of a neutrino in extra dimensions \cite{nuinED1} and then move to the general features of these models.

Consider a sterile 5D Dirac fermion $\Psi(x^{\mu},y)=(\overline{\xi}(x^{\mu},y),\eta(x^{\mu},y))^T$. 
The simplest allowed lagrangian terms include of course the 5D kinetic term and a general  brane-bulk interaction $\int d^4x \, dy \frac{h}{\sqrt{\Mf}}$ $L(x^{\mu})H(x^{\mu})\xi(x^{\mu},y)\delta(y=0) + \textrm{h.c.}$, where $h$ is a Yukawa coupling assumed to be naturally of order 1, $L=(\nu_{\ell},\ell)$ is the SM lepton doublet and $H$ the Higgs field.
In terms of a 4D description, the extra dimensional fermion is reduced to a tower of sterile fields $\psi_n$ via the expansion in Kaluza-Klein (KK) modes: $\Psi(x^{\mu},y)=\frac{1}{\sqrt{2 \pi R}}\sum_{n \in \mathbb{Z}} \psi_n(x^\mu) e^{i n \frac{y}{R}}$.
The above lagrangian terms then compose a mass matrix that involves the SM neutrino and the KK fields.
The SM neutrino ends up mixed with the sterile eigenstates with a mixing angle $\theta_n \simeq \sqrt{2}\frac{mR}{n}$, smaller and smaller as $n$ increases, and with mass gaps $(\Delta M^2)_n \simeq n^2/R^2$.
Here $m \simeq \frac{hv}{\sqrt{2 \pi R \Mf}}=\frac{hv\Mf}{M_{Pl}}\simeq10^{-4}\eV$ is a Dirac mass acquired by the neutrino.


In order to include a larger set of models in the literature \cite{nuinED2,LRRR,BCS}, we can be more general than the direct example above: what we only need to assume is that the mixing angle $\theta_n$ of the SM neutrino with the $n$-th mass eigenstate be parametrized as 
\be
\theta_n \simeq \frac{m}{\sqrt{2} M_n} \,
\ee
where now $m$ is a free parameter, not forcely related to the physical $\nu$ mass; the exact dependence of $M_n$ on $1/R$ can be a model dependent feature provided that the density of KK states is proportional to $R$.  
We allow for a separate mixing with a KK tower for each SM flavour, superimposed to the traditional flavour mixing and parametrized by the three (unrelated) quantities $m_e$, $m_\mu$ and $m_\tau$.

We require to be working in a regime of small mixing angles, to ensure the smallness of the transition probabilities and to keep under control other oscillation effects (see \cite{paper1}). This corresponds to
\be
m_e R \lesssim 10^{-5} \qquad m_{\mu,\tau} R \lesssim 10^{-4} \,.
\label{eq:mRlimits}
\ee

Equations (\ref{eq:Rlimits}) and (\ref{eq:mRlimits}) define the parameter space on the plane $m$--$R$ that we want to probe.

\subsection{Astrophysical and cosmological safety}
The framework defined above ($\Mf \sim \TeV$ and one single extra dimension under consideration) is compatible with all known constraints from astrophysics and cosmology \cite{otherconstraints}. 
Indeed, most bounds on the minimum number of extra dimensions probed by the gravitons given a certain $\Mf$ (or, conversely, on the lowest possible $\Mf$ given the number of extra dimensions) can be easily avoided if one relaxes the equality of all compactification radii.
Or one could postulate that the single extra dimension probed by sterile neutrinos is a subspace of the gravitational bulk.
On the other hand, the assumed smallness of the mixing angles is enough to protect from any undesidered drawback on BBN, CMB and so on, essentially because KK sterile neutrinos are produced through the mixing with the SM ones at a very suppressed rate in the early universe.

\section{The course of the evolution}\label{sec:evolution}
What are the modifications to the SN cooling phase described in section \ref{sec:supernova} in presence of the mixing introduced in section \ref{sec:playground}?

Let us consider a flavour eigenstate neutrino ($\nue$, $\num$ or $\nut$) produced with energy $E$ in the matter of the core by some interaction.
It immediately experiences a large MSW potential $V_{e,\mu,\tau}$ (of order of several eV) that is proportional to the local matter density and composition and it acquires an effective squared mass $m_{\textrm{eff}}^2=2EV_{e,\mu,\tau}$.
The effective mass changes along the neutrino path as the density and composition change: whenever $\meff$ equals the mass of one of the KK sterile states, a resonance occurs and the neutrino has a certain probability to {\it oscillate into a bulk sterile state, escape from the core and go lost}, carrying his energy away with him.
The same argument applies to the escape of antineutrinos if the MSW potential is negative.
The escape probability at each resonance is given by $P_n \simeq e^{-\pi\gamma_n/2}$ where the adiabaticity factor $\gamma=\genfrac{}{}{}{0}{\Delta M^2 \sin^22\theta}{2E\cos2\theta \frac{dV}{dr}/V}\ll 1$ is computable in terms of the vacuum mixing angles and $\Delta M^2$ discussed in section \ref{sec:playground}.
The disappearance probability along a distance $L$ then reads
\be
  P \left( \stackrel{\mbox{{\tiny (--)}}}{\nu}_{e,\mu,\tau} \rightarrow \textrm{bulk}\right) \simeq L\,\frac{\pi}{2\sqrt{2}}m^2_{e,\mu,\tau}R
  \left(\frac{|V_{e,\mu,\tau}|}{E}\right)^{1/2} \qquad \Big( \nu \ \textrm{if} \ V_{e,\mu,\tau}>0\,, \ \bar{\nu} \ \textrm{if} \ V_{e,\mu,\tau}<0 \Big)\, . 
  \label{eq:Ploss}
\ee
The crucial parameter $m^2R$ (with $m=m_{e,\mu,\tau}$) sets the magnitude of the escape effect and will be the subject of our analysis from now on. 
The parameter space boundaries \eq{Rlimits} and \eq{mRlimits} imply the general limits
\be
m^2_eR<10^{-8}\eV \,, \qquad  m^2_{\mu,\tau}R<10^{-5}\eV \,.
\label{eq:mmRlimits}
\ee

The escape into extra dimensions constitutes an unconventional channel for (anti)neutrino and energy loss \footnote{Notice, in passing, that matter effects play a crucial role: independently on how small the vacuum mixing angle with the $n-$th state is, the SM neutrino runs the risk of oscillating into that sterile state if the corresponding resonance is met. 
It would be not appropriate to restrict to a mixing with the lowest lying sterile states.

Also, notice that the fact that the SN core is so hot (implying $E\sim 100 \MeV$) and so dense (implying $V_{e,\mu,\tau} \sim \textrm{several} \eV$) means a large effective mass for the neutrino so that many resonances with the KK states are met. 
On the contrary, in the Sun the same mechanism is uneffective since not even the lowest resonances are met, for $1/R$ in a large portion of the range (\ref{eq:Rlimits}). 
The same is true in the case of atmospheric neutrinos. 
This protects from undesired effects due to sterile KK neutrinos in the solar and atmospheric contexts.}
that has to pass the test of the energy loss argument discussed in section \ref{sec:supernova}.\\
{\it How dangerous is this channel?}
Estimates in the literature, essentially based on the assumption of a matter potential which is constant in time, imply the very stringent bound 
$$
\mmR \lesssim 10^{-12}\eV, 
$$
cutting a large portion of the ranges in (\ref{eq:mmRlimits}).\cite{BCS}
Within such a limit, the escape process is too small to have any effect and the SN evolution is completely untouched.
We want to reconsider this conclusions in more details.

\subsection{The feedback mechanism}
\begin{figure}
\begin{center}
\epsfig{file=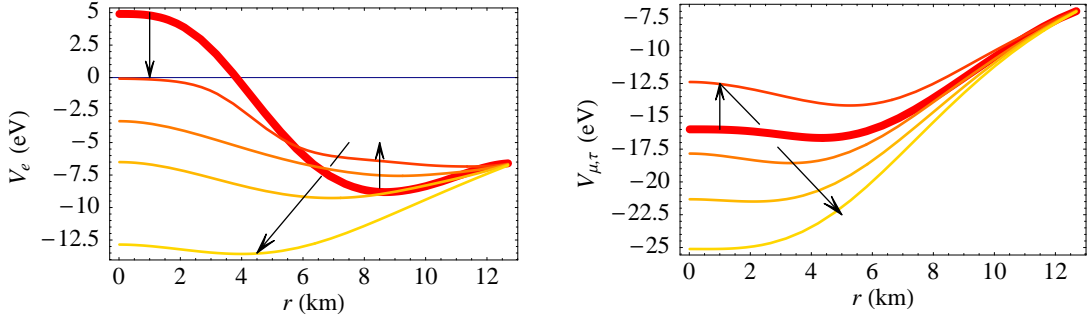,width=0.90\textwidth}
  \caption{Profile of the MSW potentials experienced by electron neutrinos (left) and muon and tau neutrinos (right). The thick line is a typical initial configuration; the thin lines are snapshots at 1, 5, 10 and 20 secs, for $m_e^2 R=10^{-9}\eV$ on the left and $m_{\mu,\tau}^2 R=10^{-7}\eV$ on the right.}
\label{fig:potentials} 
\end{center}
\end{figure}

The matter potentials probed by electron and non-electron neutrinos are respectively~\footnote{In general, a $Y_{\nu_{\mu,\tau}}$ term could be present in $V_e$, but it never moves significantly from zero in the case of extra dimension open only to electron neutrinos considered below.}
\be
 V_e = \sqrt{2}\GF\nB\left(\frac{3}{2}Y_e+2Y_{\nu_e}-\frac{1}{2}\right) \quad V_{\mu,\tau} = \sqrt{2}\GF\nB\left(\frac{1}{2}Y_e+2Y_{\nu_{\mu,\tau}}+Y_{\nue}-\frac{1}{2}\right) \, .
\label{eq:potentials}
\ee
Their initial configurations are the darkest curves depicted in \fig{potentials}.
Let us first consider the case where the extra dimension is open for the electron flavour and focus on the potential $V_e$. 
In the region where $V_e>0$, $\nue$ quickly escape into the bulk, thus reducing their contribution $\YLe$ to the potential, that is then pushed to zero.\cite{LRRR} 
Since the escape probability is proportional to $V_e$, this stops the escape itself.
Similarly, in the region where $V_e<0$, $\nueb$ escape, thus increasing $\YLe$ and forcing the potential towards zero.
Subsequently, neutrino diffusion starts to deplete the relative fractions $Y_x$ and pulls $V_e$ below zero.
Again, then, $\nueb$ escape tends to pull $V_e$ back to zero. 
A non trivial feedback mechanism on the escape process is thus at work.\cite{paper1}

In the case that the extra dimension is seen by the muon (or tau) flavour, the potential $V_{\mu,(\tau)}$ is negative everywhere so that $\bar{\nu}_{\mu,(\tau)}$ escape into the bulk.
This generates a positive $Y_{\num,(\tau)}$, the balance $\nu$--$\bar{\nu}$ is broken and a positive chemical potential arises.  
This in turn inhibits the escape itself, both because $V_{\mu,(\tau)}$ is lifted towards zero by the term $Y_{\nu_{\mu,(\tau)}}$ in \eq{potentials} and, more important, because the $\bar{\nu}_{\mu,(\tau)}$ abundance is suppressed in the presence of the chemical potential.
A feedback mechanism on the escape process is again at work.\cite{paper2}

To study the above picture quantitatively, we use a simplified model of the SN core dynamics\cite{paper1} that essentially superimposes to the usual neutrino diffusion the escape effect.
We consider the cases of extra dimension open for $\nue$, $\num$ or $\nut$ one at a time, each for several values of the corresponding extra dimensional parameter $m^2_{e,\mu,\tau}R$. 

\vspace{-0.2cm}
\section{The outcome}\label{sec:outcome}
\begin{figure}
\begin{center}
\epsfig{file=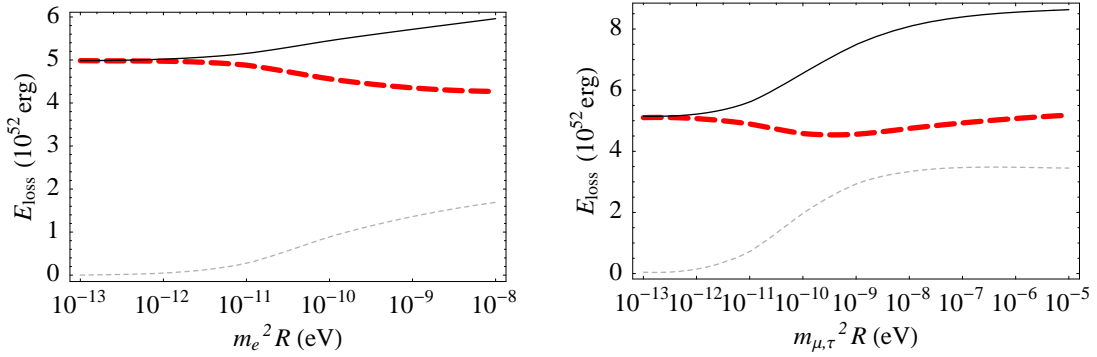,width=0.90\textwidth}
  \caption{Energy that leaves the SN core in the first 10 secs as a function of the parameter $\mmR$, for the case of extra dimension open to electron neutrinos (left) or muon or tau neutrinos (right). The portion emitted in usual visible neutrinos is highlighted (red dashed line). See also shown the total energy (black solid line) and the portion carried into the bulk by sterile neutrinos (lower dotted line).}
\label{fig:energies}
\vspace{-0.0cm} 
\end{center}
\end{figure}

\begin{figure}[t]
\begin{center}
\epsfig{file=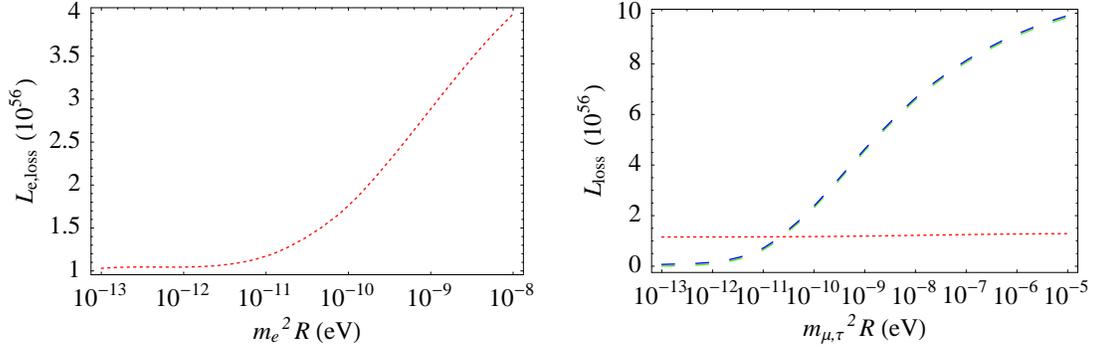,width=0.90\textwidth}
  \caption{Lepton numbers emitted from the SN core in the first 10 secs as a function of $\mmR$. Left: the electron lepton number emitted in the case of extra dimension open to electron neutrinos. Right: the muon or tau lepton number emitted in the case of extra dimension open to muon or tau neutrinos (blue-green dashed line) and electron lepton number simultaneously emitted (red dotted line).}
\label{fig:Lnumbers}
\vspace{-0.0cm} 
\end{center}
\end{figure}

\vspace{-0.3cm}
Let us analyse the relevant outputs of the modified core evolutions as a function of $\mmR$.\\
First, we observe in \fig{energies} that the reduction of the portion of energy emitted into the visible channel is always limited to a $\sim20\%$ of the standard ($\mmR=0$) case, for all escape scenarios. This is thanks to the action of the feedback mechanism.
Second, one also checks, by inspecting the behaviour of the profiles of the dynamical variables, that the characteristic timescale of the core evolution (and thus of the neutrino emission) is only marginally affected.\\
We can therefore conclude that the {\it energy loss constraint} discussed in section \ref{sec:supernova} is {\it passed}: no direct upper bounds on the parameters $m^2_{e,\mu,\tau}R$ need to be imposed. 
They only remain subject to the generic bounds in (\ref{eq:mmRlimits}).\\
The other significant result (see \fig{Lnumbers}) is the {\it emission of an excess of net lepton number} with respect to the standard case, counterbalancing the antineutrinos that have escaped into the bulk.
This will have interesting phenomenological consequences on the $\nu$ signal observable on Earth.

\vspace{-0.2cm}
\section{The signatures}\label{sec:signatures}
\begin{figure}[t]
\begin{center}
\vspace{-0.5cm}
\psfig{figure=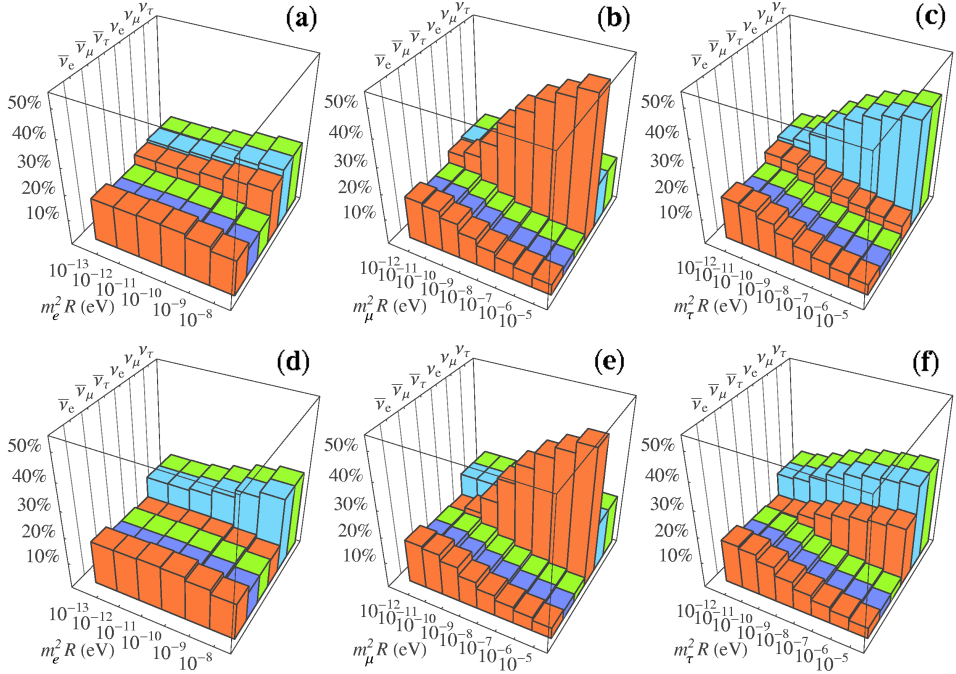,height=10cm}
\caption{Indicative composition of the neutrino flux reaching the Earth's surface, as a function of $\mmR$, for the different cases of escape into the bulk: $\nue$ (first column), $\num$ (second column), $\nut$ (third column); for $\sin^2\theta_{13}<10^{-5}$ (first line) and for $\sin^2\theta_{13}>10^{-3}$ (second line).
\label{fig:finalcomposition}}
\vspace{-0.5cm}
\end{center}
\end{figure}

\vspace{-0.3cm}
The neutrinos and antineutrinos emitted from the core are subject to several vicissitudes before they constitute the signal that can be detected on Earth. 
First is the energy redistribution inside the neutrino-spheres, then they undergo matter flavour oscillations in the peripheric low density region of the star and vacuum flavour oscillations in the journey from the supernova to Earth.
The indicative percentual composition of the neutrino signal on Earth as a function of $\mmR$ is represented in \fig{finalcomposition}, assuming present global best fit values for $\theta_{12}$, $\theta_{23}$, $\Delta M^2_{12}$ and $|\Delta M^2_{23}|$, taking into account the current upper bound on $\theta_{13}$ and choosing e.g. the case of normal hierarchy.

Although the compositions are quite case-dependent, some structures can be observed.\\
First of all, the flux ratio of neutrinos over antineutrinos (independently on their flavours) is always enhanced.
This is a general feature of supernova neutrino oscillations into extra dimensions that can be traced back to the escape of antineutrinos into the bulk due to the predominant negativity of matter potentials.
In cases (a), (b), (e) and (f) the enhancement shows in the $\nue$/$\nueb$ ratio, that has better chances of future observation.
In cases (c) and (d) one has to look at the flux of $\nu_{\mu,\tau}$ and $\bar{\nu}_{\mu,\tau}$.\\
Second, the $\nueb$ flux is generally reduced, although never more than $\sim 60\%$.
This guarantees the compatibility with the $\nueb$ signal in \SN, also given its low statistics.\\
Finally, an indication on which flavour oscillated into the bulk can come from the $\nue$ spectrum.
Indeed, if an excess of muon or tau neutrinos is emitted from the core, this produces a harder $\nue$ spectrum when they pour into $\nue$ due to the game of flavour oscillations, since they are more energetic.
The contrary is true in the case of electron neutrinos in excess.

\vspace{-0.2cm}
\section{Conclusion}\label{sec:conclusions}
\vspace{-0.3cm}
We discussed the mixing of SM neutrinos with bulk sterile fermions, a natural and general feature of models with extra dimensions. 
This mixing provides an unconventional escape channel for neutrinos in the
supernova core during the cooling phase, which could in principle give
strong bounds on the parameters of the extra dimensions when facing
the observed \SN\ signal.
The process mainly consists in the escape of antineutrinos into the bulk, acting in parallel with diffusion.  
We showed that a feedback mechanism turns on, preventing
extreme and unacceptable modifications of the evolution to occur so that no direct bound need to be imposed and a large portion of the parameter space is regained for extra dimensions.\\
In that portion, although in such a `safe' way, the supernova core evolution is
however affected, and we discussed the main consequences.
The compatibility with the \SN\ signal is preserved once one is willing to
accept a certain reduction of the $\nueb$ component. 
We indicated as signatures, in cases of future galactic SN events, a general
dominance of neutrino flux over antineutrinos and, depending on the
flavour that mixes with bulk fields, some peculiar structures of the
relative enhancements in the neutrino fluxes and a harder or softer
$\nue$ spectrum.

\begin{footnotesize}
\vspace{-0.2cm}
\section*{Acknowledgments}
I warmly thank Andrea Romanino, Giacomo Cacciapaglia and Lin Yin for the precious collaboration. It's also a pleasure to thank the Organizing Committee of the Moriond conference.


\end{footnotesize}

\end{document}